\documentstyle[epsf,prl,aps,twocolumn]{revtex}
\def\be{\begin{equation}}
\def\ee{\end{equation}}
\def\bea{\begin{eqnarray}}
\def\eea{\end{eqnarray}}

\def\rd{\mbox{d}}

\def\t12h{\frac{\theta_{12}}{2}}
\def\f12{\frac{1}{2}}

\def\eps{\varepsilon}

\def\r#1{(\ref{#1})}
\def\nn{\nonumber\\}

\def\NPB#1#2#3{{\sl Nucl. Phys.} {\bf B#1} (#2) #3}

\def\PRB#1#2#3{{\sl Phys. Rev.} {\bf B#1} (#2) #3}

\begin {document}
\draft
\title{\bf Optical conductivity of one-dimensional Mott insulators
}
\author {D. Controzzi$^{(a)}$, F. H. L. Essler$^{(b)}$ and
 A. M. Tsvelik$^{(a)}$
}
\address{$^{(a)}$ Department of Physics, 
        University of Oxford, 1 Keble Road, Oxford OX1 3NP, UK}
\address{$^{(b)}$ Department of Physics, Warwick University, 
Coventry CV4 7AL, UK}

\maketitle

\begin{abstract}
We calculate the optical conductivity of one-dimensional Mott
insulators at low energies using a field theory description. The
square root singularity at the optical gap, characteristic of band
insulators, is generally absent and appears only at the Luther-Emery
point. We also show that only few particle processes contribute
significantly to the optical conductivity over a wide range of
frequencies and that the bare perturbative regime is recovered only at
very large energies. We discuss possible applications of our
results to quasi one-dimensional organic conductors.
\end{abstract}

\narrowtext
Measurements of dynamical properties and in particular the optical
conductivity $\sigma(\omega)$ are supposed to provide a stringent test
of the existing theories of quasi one-dimensional (1D) systems. The
behaviour of $\sigma(\omega)$ in the metallic regime is easily
understood in terms of the Tomonaga-Luttinger theory \cite{LL}. The
situation in the Mott insulating phase \cite{mott} is much more
complicated as a spectral gap is dynamically generated by
interactions. 
Here $\sigma(\omega)$ has until now only been studied by perturbative
methods \cite{gia1,giamarchi}, which are expected to work well at high
and intermediate frequencies but are not applicable to the most
interesting regime of frequencies close to the optical gap.
The purpose of the present work is to determine $\sigma(\omega)$ in 1D
Mott insulators for all frequencies much smaller than the bandwidth,
which is the large scale in the field theory approach to the problem.
In particular we obtain for the first time the true behaviour of 
$\sigma(\omega)$ just above the optical gap.

An important property of one-dimensional systems that significantly
simplifies our analysis is spin-charge separation, which occurs at 
energies much smaller than the bandwidth. 
In this regime $\sigma(\omega)$ is determined solely by
the charge degrees of freedom. The standard description of the charge
sector of 1D Mott insulator is given by the sine-Gordon model (SGM)
\cite{elp,giamarchi} 
\be
H_{sG}=\int \rd x 
\left [ 4\pi( \Pi )^2+ \frac{1}{16 \pi }
(\partial _x  \phi)^2  +  2\mu\cos( \beta \phi)\right].
\label{ll}
\ee
Here the momentum and coordinate densities obey the standard 
commutation relation $[\Pi(x), \phi(y)] = - i\delta(x -
y)$. Throughout this letter we set the charge velocity and
$\hbar$ equal to one. 

The cosine term in the Hamiltonian is related to Umklapp processes and
the value of the sine-Gordon coupling constant $\beta$ is determined by
the interactions. The Umklapp processes are relevant for $\beta^2 < 1$
and dynamically generate a spectral gap $M$, which is related to $\mu$
by (\ref{gap}). 
For $1/2 < \beta^2 < 1$ the spectral gap is related to optical gap $\Delta$
(i.e. the gap seen in the optical absorption) by $\Delta=2M$ whereas
for $\beta^2<1/2$ solitonic bound states are formed below $2M$.

Our calculations of $\sigma(\omega)$ are based on the exact solution 
of the SGM and in particular on the work of Smirnov \cite{smirnov}. We
confine our analysis to the repulsive regime  $1/2 < \beta^2 < 1$,
where the excitation spectrum consists of charged particles and holes
(solitons and anti-solitons), which do not form bound states.
At the ``Luther-Emery'' point $\beta^2 = 1/2$ the SGM is equivalent to
the theory of free spinless massive Dirac fermions. In this limit the
solitons become non-interacting particles and the Mott insulator turns
into a conventional band insulator.
In the limit $\beta^2 \rightarrow 1$ the SGM acquires an SU(2) 
symmetry and describes the Hubbard model at half-filling in
the regime of weak interactions \cite{affleck89,smat} and $\sigma(\omega)$ 
was recently determined in \cite{hubb}.

The optical conductivity is related to the imaginary  part of the
current-current correlation function, $\chi(\omega,q)=
\langle j_{-q} \; j_q\rangle$ by
\be
\label{sigma}
\sigma(\omega >0) = {\rm Im}
\left \{ \chi(\omega,q=0) \right \} /\omega.
\ee
The current density operator is proportional to the momentum density 
\be
j_q = A^{1/2} \Pi_q, ~~ \Pi_q = \int dx \Pi(t, x)e^{iqx}\ .
\ee
The non-universal coefficient $A^{1/2}$ depends on the detailed
structure of the underlying microscopic lattice model.

Using the spectral representation one can express the optical
conductivity at $T = 0$ as a sum over matrix elements of the zero 
wave vector  Fourier component of the momentum operator:
\be
\sigma(\omega >0) = \frac{A}{\omega}\sum_n |\langle
0|\Pi_0|n\rangle|^2\delta[ \omega-(E_n-E_0)]. 
\label{chi}
\ee
Here  $|0\rangle$ and $|n\rangle$ represent the ground state and 
excited states with energies $E_0$ and $E_n$ respectively. 
The difficulties in computation of the optical response
are related to the fact that one  requires  not only the knowledge of
the spectrum $E_n$, but also of the matrix elements of the momentum operator. 
The exact expressions for the matrix elements $\langle n|\Pi_0|0\rangle$ are 
extracted from the exact solution by means of the so-called form
factor bootstrap procedure \cite{smirnov}. This approach is
particularly efficient for strongly interacting integrable models with
spectral gaps, because for a given energy $\omega$ the spectral representation
for the imaginary part contains only a {\it finite} number of
terms (in the absence of bound states at most $[\omega/\Delta]$ terms). In 
practice the spectral sum is found to converge extremely rapidly, so
that a very good approximate description can be obtained by taking
into account intermediate states with at most four particles
\cite{oldff}. The multi-particle matrix elements become essential only
at very high energies where the field theory can no longer be used to
describe the underlying lattice model anyway.

In order to compute (\ref{chi}) we need to introduce a suitable
spectral representation. In the parameter regime we study,
the spectrum contains only solitons and anti-solitons with
relativistic dispersion $e(p)=\sqrt{p^2+M ^2}$. It is useful to
parametrize the spectrum in terms of a rapidity variable $\theta$
\be
p = M\sinh\theta, \: e = M\cosh\theta\ ,\qquad
\ee
Solitons and anti-solitons are distinguished by the internal index
$\eps = \pm$.
A state of $n$ solitons/anti-solitons with rapidities $\{
\theta_k \}$ and internal indices $\{\eps_k\}$ is denoted by: $
|\theta_n\ldots\theta_1\rangle_{\eps_n\ldots\eps_1} 
\label{states}$.
Its total energy $E$, momentum $P$ and electric charge $Q$ are:  
\be
P = M\sum_{k=1}^n\sinh\theta_k,\ E= M \sum_{k=1,}^n \cosh\theta_k,\
Q \propto \sum_{k=1}^n \eps_k \ .
\ee 
In terms of this basis $\sigma(\omega)$ is expressed as
\bea
\sigma(\omega)&=& \frac{2 \pi^2A}{\omega}
\sum_{n=0}^\infty\sum_{\eps_i}\int 
\frac{\rd\theta_1\ldots\rd\theta_n}{(2\pi)^n n!} \left |
f^j(\theta_1\ldots\theta_n)_{\eps_1\ldots\eps_n}
\right | ^2 \nn
 &\times&\delta(M \sum_k\sinh\theta_k)\delta(\omega - M \sum_k
\cosh\theta_k) \nn
&=&\sigma_2(\omega)+\sigma_4(\omega)+ ...
\label{expansion2}
\eea

Here 
\be
\label{ff}
f^{j}(\theta_1\ldots\theta_n)_{\eps_1\ldots\eps_n}\equiv
\langle 0| j(0,0)|\theta_n\ldots\theta_1\rangle_{\eps_n\ldots\eps_1}
\ee
are the form factors of the current operator, 
$\sigma_2(\omega)$ and $\sigma_4(\omega)$ represent the
contributions from 2 and 4-particle processes and the dots
indicate processes involving higher number of (anti)solitons. 
We note that as a consequence of symmetry properties only 
intermediate states with an even number of particles contribute 
to this correlation function. From (\ref{expansion2}) it is easy to
see that only 2-particle processes contribute up to energies
$\omega=4 M$, only 2 and 4-particle processes up to $\omega=6 M$ and
so on. 

The form factors (\ref{ff}) have been determined in \cite{smirnov}
and can be used to calculate the first few terms in the
expansion (\ref{expansion2}). We find
\be
\sigma_2(\omega)= \frac{2A\Theta(\omega-2M)}{{\omega^2}
\sqrt{{\omega}^2-4M^2}}|f(\theta)|^2 \ ,\label{low}
\ee
where $\Theta(x)$ is the Heaviside function,
\bea
&&f(\theta)=f^j(\theta)_{+-}=f^j(\theta)_{-+}=\frac{2\pi M}{i\beta}
\frac{\sinh{\theta/2}}{\cosh{ \left ( \frac{\theta+i \pi}{2 \xi}
\right )}} \nn
&&\times\exp \left \{ \int_0^\infty dt \; \frac{\sinh^2{t(1-i
\theta/\pi)} \sinh{t(\xi-1)}}{t \sinh{2 t} \cosh{t} \sinh{t\xi}}
\right\}
\eea
and 
\be
\theta= 2 {\rm arccosh(\tilde{\omega})}\ ,\ 
\xi=\beta^2/(1-\beta^2)\ ,\ \tilde{\omega}=\omega/2M.
\ee

The four particle contribution is of the form
\bea 
\label{four}
&& \sigma_4(\omega)=\frac{\Theta(\omega-4M)}{192\omega \pi^2 M^2} \sum_{\eps_i}
\sum_{\sigma=\pm}
\int_{-a}^{a}
 d\theta \; \int_{-b(\theta)}^{b(\theta)} d \gamma
\nn
&&\times|f^j(g-\frac{\sigma\alpha}{2},g+\frac{\sigma\alpha}{2}
g+\theta+\gamma,
g-\theta+\gamma)_{\eps_1...\eps_4}|^2
\nn
&&\times\left\{\left(
\sqrt{\cosh^2{\theta}\sinh^2{\gamma}+\tilde{\omega}^2}
-\cosh{\theta}\cosh{\gamma}
\right)^2-1\right\}^{-\frac{1}{2}}\nn
&&\times\left[\cosh^2{\theta}\sinh^2{\gamma}+\tilde{\omega}^2\right]^{-\frac{1}{2}}
\ ,
\eea
where 
\bea
a&=&{\rm arccosh(\tilde{\omega}-1)}, \ 
b(\theta)={\rm arccosh\left[\frac{\tilde{\omega}^2-1-\cosh^2{\theta}}
{2 \cosh{\theta}}\right]}, \nn
g&=&\ln\left[\frac{\cosh(\alpha/2)+\exp(-\gamma)\cosh\theta}
{\tilde{\omega}}\right], \nn
\alpha&=&2 {\rm arccosh}\left[
\sqrt{\cosh^2{\theta}\sinh^2{\gamma}+\tilde{\omega}^2}- 
\cosh{\theta}\cosh{\gamma}\right]. \nonumber 
\eea
The four particle form factor is given by
\bea
\label{fo}
&&f^j(\beta_1,...,\beta_4)_{--++}
=\frac{4 \pi^3}{\beta}\xi M d^2 \prod_{k<l}
\zeta(\beta_k-\beta_l) \nn
\times&&\prod_{m,n=1,2}\left({\sinh[(\beta_{2+m}-\beta_n-i \pi)/\xi]}
\right)^{-1} \nn
\times&&2 \sinh[{(\beta_4+\beta_3-\beta_1-\beta_2-2 \pi i)/2 \xi}]
\nn
\times&&\exp(-\frac{1}{\xi} \sum_k \beta_k)  
\int \frac{\rd \alpha}{i \pi}\prod_k
\varphi(\alpha-\beta_k) \cosh(\alpha-\frac{1}{2}\sum_k \beta_k) \nn
\times&&\Delta(e^{2 \alpha/\xi}|e^{2 \beta_1/\xi},
e^{2 \beta_2/\xi}|e^{2 \beta_3/\xi},e^{2 \beta_4/\xi}).
\eea
The different orderings for (\ref{fo}) that appear in
(\ref{four}) can be obtained using the following property of the form
factors
\bea
&&f^j(\theta_1\ldots \theta_i\theta_{i+1}\ldots
\theta_n)_{\eps_1\ldots \eps_i\eps_{i+1}\ldots\eps_n}
S^{\eps_i\eps_{i+1}}_{\eps_i'\eps_{i+1}'}(\theta_i-\theta_{i+1})\nn
&&= f^j(\theta_1\ldots \theta_{i+1}\theta_{i}\ldots
\theta_n)_{\eps_1\ldots \eps_{i+1}'\eps_{i}'\ldots\eps_n}\ ,
\label{sym}
\eea
where $S$ is the two body scattering matrix.
The various functions appearing in Eqs (\ref{fo}) and (\ref{sym})
can be found in \cite{smirnov} (note that our definition of $\xi$ 
differs by a factor of $\pi$).

\begin{figure}[ht]
\begin{center}
\epsfxsize=0.45\textwidth
\epsfbox{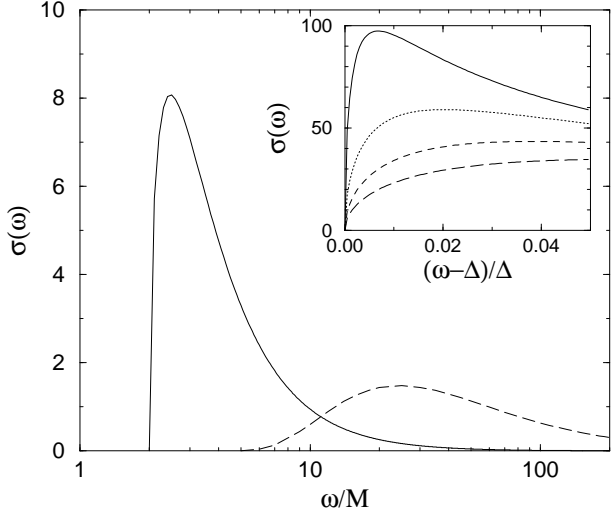}
\end{center}
\caption{\label{fig:2p4}
Two particle (solind line) and one hundred times the four-particle
(dashed line) contributions to the optical conductivity as a function
of $(\omega/M)$ for $\beta^2=0.9$.  Inset: threshold behaviour of the
optical conductivity close to the Luther-Emergy point for four
different values of $\beta$; $\beta=0.75$
(solid), $\beta=0.74$ (dotted), $\beta=0.73$ (dashed) and $\beta=0.72$
(long dashed) .
}
\end{figure}

The two and (one hundred times the) four-particle contributions to
$\sigma(\omega)$ for $\beta^2=0.9$ are presented in Fig.1. Most
importantly, the square root singularity, being a characteristic
feature of band insulators, is suppressed by the momentum dependence
of the soliton-antisoliton form factor and reappears only for the
Luther-Emery point $\beta^2 =1/2$.  
This effect was noted previously for the Hubbard model
at half-filling \cite{hubb} which corresponds to the special
SU(2)-symmetric point $\beta^2 = 1$. We find that for {\sl any}
$\beta^2\neq 1/2$ there is a square root ``shoulder''
$\sigma(\omega)\propto\sqrt{\omega-\Delta}$ for $\omega/\Delta-1\ll 1$
as is shown in the inset of Fig.1.
In the vicinity of the Luther-Emery point $\beta^2 =1/2$ we obtain the
following analytical expression valid for ${\tilde\omega} - 1 \ll 1$: 
\begin{eqnarray}
\sigma(\omega) \propto \frac{\sqrt{\tilde\omega^2 - 1}}{[\tilde\omega^2
-1] + \xi^2\sin^2\gamma},~~ \gamma = \pi\left(\frac{1}{2\beta^2} -
1\right).
\end{eqnarray} 
The square root singularity above $\omega=\Delta$ for $\beta^2=1/2$ is
replaced by a maximum occurring at $\omega/\Delta - 1 \propto \gamma^2$.

The four particle contribution to $\sigma$ is seen to be insignificant
at low energies and becomes larger than the two particle contribution
only at $\omega \approx 180 M$ for $\beta^2=0.9$. 
This suggests that the optical conductivity is well described by the 
combination of 2 and 4-particle contributions up to several hundred 
times the mass gap.
Computation of higher order terms in Eq.(\ref{expansion2}) becomes
cumbersome and probably of no physical interest, since the previous
analysis suggests that they become important outside the region of
applicability of the field theory approach to physical
systems.

At frequencies much larger than the gap it is possible to determine
$\sigma(\omega)$ by perturbative methods. The leading asymptotics
can be calculated by ``conformal perturbation theory''
\cite{cpt}. Here the cosine interaction in (\ref{ll}) is considered as
a (relevant) perturbation of the Gaussian model and correlation
functions are calculated in a perturbative expansion in powers of the
scale $\mu$, which then can be expressed in terms of the physical gap
$M$ as \cite{abz} 
\be
\mu=\frac{\Gamma(\beta^2)}{\pi \Gamma(1-\beta^2)}\left [
M\frac{\sqrt{\pi}
\Gamma(1/2+\xi/2)}{2 \Gamma(\xi/2)} \right ]^{2-2 \beta ^2}.
\label{gap}
\ee
We find to leading order
\bea
&&\sigma(\omega)=2^{9-4\beta^2}\left(\frac{\pi^2\beta}
{\Gamma(2\beta^2)}\right)^2\mu^2 \omega^{(4 \beta^2-5)}\nn
&&=\frac{8\pi^3\beta^2}{\omega\Gamma^2(1-\beta^2)\Gamma^2(\frac{1}{2}+\beta^2)}
\left[\frac{\Gamma(\frac{\xi}{2})}{2\sqrt{\pi}\Gamma(\frac{1+\xi}{2})}
\frac{\omega}{M}\right]^{4\beta^2-4}.
\label{high}
\eea
We emphasize that the ratio of the coefficients of the high- and
low-energy asymptotics \r{high}, \r{low} is {\sl fixed}
\cite{smirnov},\cite{lukyanov}. In other words, the amplitude of the
power law in \r{high} is tied to the overall factor in \r{low} and
the form factor expansion must approach the perturbative result in the
large-$\omega$ limit. 
A comparison between the form factor results and \r{high} is shown in Fig.2.
We see that the asymptotic regime is not yet reached at energies as
high as $\omega \sim 1000 M$ (in practical terms this implies that
perturbation theory cannot be used to make contact with experiment).
We note that the contributions due to intermediate states with 6,8,10
... particles are all positive and will make the agreement of the form
factor sum with perturbation theory in the region  $\omega\approx 1000M$
only worse.
A good way to overcome these deficiencies of bare
perturbation theory is to carry out a renormalisation-group (RG)
improvement as performed in \cite{gia1}. In Zamolodchikov's scheme
\cite{abz} the RG equations for the Sine-Gordon model are given by
\be
\frac{dg_\perp}{dt}=\frac{g_\parallel g_\perp}{1+\frac{g_\parallel}{2}}\ ,\quad
\frac{dg_\parallel}{dt}=\frac{g^2_\perp}{1+\frac{g_\parallel}{2}}\ .
\label{rgeq}
\ee
The solution of \r{rgeq} is
\be
g_\perp=4\frac{1-\beta^2}{\beta^2}\frac{\sqrt{q}}{1-q}\ ,\quad
g_\parallel=2\frac{1-\beta^2}{\beta^2}\frac{1+q}{1-q}\ ,
\label{rg1}
\ee
where
\be
q\left(\frac{(1-q)\beta^2}{4(1-\beta^2)}\right)^{2\beta^2-2}=e^{(4-4\beta^2)
(t-t_0)}.
\label{rg2}
\ee
Using $t-t_0=\ln\left(\frac{\sqrt{\pi}e^{3/4}M}{2^{3/2}\omega}\right)$
we can reexpress \r{high} up to higher order terms as
\be
\sigma(\omega)=\frac{\pi^3\beta^6
g_\perp^2}{2\omega\Gamma^2(2-\beta^2)\Gamma^2(\frac{1}{2}+\beta^2)} 
\left[\frac{\Gamma(\frac{\xi}{2})e^{3/4}\sqrt{\xi}}{2^{7/2}\Gamma(\frac{1+\xi}{2})}
\right]^{4\beta^2-4}.
\label{rg}
\ee
The RG improved result \r{rg} for $\sigma(\omega)$ is compared to the
form factor result (sum of the two and four-particle contributions) in
the inset of Fig.2. The agreement is rather good down to energies of
the order of $5M$.

\begin{figure}[ht]
\begin{center}
\epsfxsize=0.45\textwidth
\epsfbox{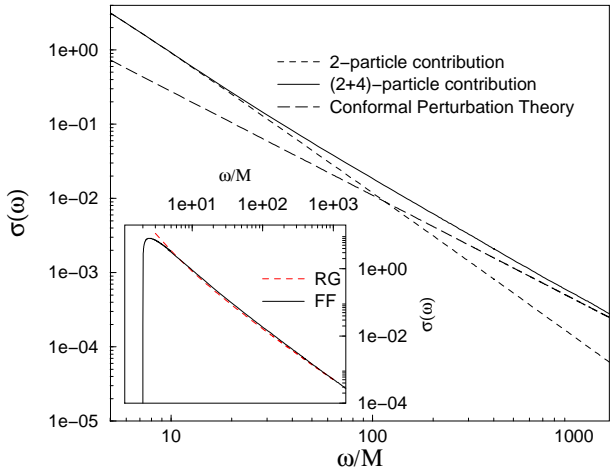}
\end{center}
\caption{\label{cpt.fig}
Comparison between the 2 and 2+4-particle contribution to the optical
conductivity and the perturbative result, for $\beta ^2=0.9 $. Inset:
comparison between the form factor result and RG improved perturbation
theory.
}
\end{figure}

One possible realisation of a 1D Mott insulator are the $\rm (TMTSF)_2X$
Bechgaard salts \cite{exp}. These materials are highly
anisotropic and can be modelled as weakly coupled, quarter-filled
chains. At energies or temperatures above the 1D-3D crossover scale
$E_{\rm cr}$ the interchain coupling becomes ineffective and a
description in terms of a purely 1D model with charge sector \r{ll}
should be possible \cite{giamarchi}. At present there is some
uncertainty regarding the value of $E_{\rm cr}$ because interactions
can renormalize its bare value, set by the interchain coupling,
downwards \cite{boies}. 
There is a lot of ambiguity in fitting our results to the data. The
value of the optical gap $2M$ is not known and, as discussed above, we
cannot calculate the overall normalisation of $\sigma(\omega)$. We
therefore use these as parameters in order to obtain a good fit at
large $\omega$ (where the theory is expected to work best as 3D
effects are unimportant) to the data \cite{exp} for any given value of
$\beta$. We obtain reasonable agreement
with the data for $\beta^2\approx 0.9$, which corresponds to a Luttinger 
liquid parameter of $K_\rho\approx 0.23$. This value is consistent with 
previous estimates (see the discussion in \cite{exp}). 
\begin{figure}[ht]
\begin{center}
\epsfxsize=0.45\textwidth
\epsfbox{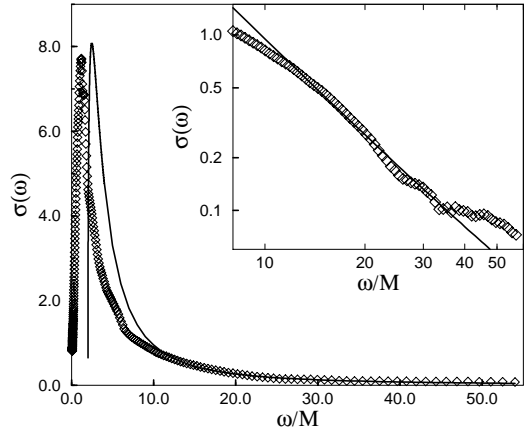}
\end{center}
\caption{
Comparison between the optical conductivity  calculated in the
SGM for $\beta^2=0.9$ (solid lines) and measured optical
conductivity for $\rm (TMTSF)_2PF_6$ from Ref.[12] (diamonds). The
inset shows the same comparison on a logarithmic scale.}
\end{figure}

As is clear from Fig. 3, the model \r{ll} seems to apply well at high 
energies, but becomes inadequate at energies of the order of about $10$ 
times the Mott gap ($\approx 1600/{\rm cm}$ in $\rm (TMTSF)_2PF_6$). Spectral
weight is transferred to lower energies and physics beyond that of a pure 1D
Mott insulator emerges. There are at least two mechanisms that should be
taken into account in this range of energies. Firstly, a small
dimerization occurs in the 1D chains and will almost certainly affect
the structure of $\sigma(\omega)$ around its maximum. Secondly, the
interchain hopping is no longer negligible \cite{interchain} and ought
to be taken into account.

In summary, we have exactly calculated $\sigma(\omega)$ for a pure 1D Mott
insulator in a low-energy effective field theory approach. We have
determined the threshold behaviour for the first time and found it to
exhibit a universal square root increase for any $\beta^2>1/2$.
This is in contrast to the well-known suqare-root singularity that
appears at the Luther-Emery point $\beta^2=1/2$.
In the ``low'' energy region ($\omega/\Delta <50$) the optical
conductivity is dominated by the two-particle form factor contribution
with a small correction from four-particle processes. This means that
the entire optical transport is dominated by two-particle processes!
We furthermore have shown that the leading asymptotic behaviour
obtained in perturbation theory is a good approximation only at
extremely large frequencies, whereas RG-improved perturbation theory
works well over a large region of energies.

We are grateful to A. Schwartz for generously providing us with the
experimantal data and to F. Gebhard, T. Giamarchi, E. Jeckelmann and
S. Lukyanov for important comments and discussions. We thank the
Isaac Newton Institute for Mathematical Sciences, where this work was
completed, for hospitality.

\narrowtext
\end{document}